# A test of Einstein's theory of gravitation: Velocity distribution of low-energy particles in a spherically symmetric gravitational field

Jian-Miin Liu*

Department of Physics, Nanjing University

Nanjing, The People's Republic of China

*On leave. E-mail address: liu@phys.uri.edu.

**Abstract**

We propose a new test of Einstein's theory of gravitation. It concerns the velocity distribution of low-energy particles in a spherically symmetric gravitational field.

**Introduction.** Gravity is a fundamental interaction of Nature. Differing all other interactions, it possesses a specific property that the movement of an object under gravitational interaction is independent of the mass of the object [1-5]. This property leaded to Einstein's Riemannian-geometric description of gravity. Einstein's theory of gravitation has held up in several experimental tests such as the gravitational red shift, the deflection of light in a gravitational field, the perihelion advance of planetary orbits, the delay of radar pulses in a gravitational field [3,4].

In this letter, we propose a new test. It concerns the velocity distribution of low-energy particles in a spherically symmetric gravitational field.

Suppose we have N identical classical particles of low energy in a relatively small volume located at point $(x^1,x^2,x^3)$ in a spherically symmetric gravitational field produced by mass M. Our question is: What is the velocity distribution of these N particles?

**Velocity space in a spherically symmetric gravitational field.** The Schwarzschild line element in the spherically symmetric gravitational field of mass M, according to Einstein's theory of gravitation, is [4]

$$d\Sigma^2 = g_{ij}(x)dx^i dx^j, \quad i,j=1,2,3,4, \tag{1a}$$

$$g_{rs}(x) = \delta_{rs} + \frac{r_s/r}{1 - r_s/r} \frac{x^r x^s}{r^2}, \quad r,s=1,2,3, \tag{1b}$$

$$g_{r4}(x) = g_{4r}(x) = 0, \tag{1c}$$

$$g_{44}(x) = (1 - \frac{r_s}{r}), \tag{1d}$$

$$r_s = 2GM/c^2, \tag{1e}$$

where $x^4 = ict$, denominator $r = [(x^1)^2 + (x^2)^2 + (x^3)^2]^{1/2}$, c is the speed of light and G is the Newton gravitational constant. This line element can be written in the 3X1 form,

$$dX^2 = g_{rs}(x)dx^r dx^s, \quad r,s=1,2,3, \tag{2a}$$

$$dT^2 = (1-\frac{r_s}{r})dt^2. \tag{2b}$$

Here dX and dT are respectively the actual differential distance and time interval

In the presence of the spherically symmetric gravitational field, the local structures of space and time are no longer of the Euclidean. When and only when M=0, they are the Euclidean. For non-zero M, the local structures of space and time are Euclidean only at r=∞, far from the mass M. From Eqs.(2a-2b), we find

$$Y^2 = \beta^2 g_{rs}(x) y^r y^s, \quad r,s=1,2,3, \tag{3}$$

with

$$\beta = 1/(1-r_s/r)^{1/2}, \tag{4}$$

where $y^r = dx^r/dt$, r=1,2,3, is the Newtonian velocity, Y=dX/dT is the actual speed or velocity-length. Eq.(3) embodies the velocity space at point $(x^1, x^2, x^3)$,

$$dY^2 = \beta^2 g_{rs}(x) dy^r dy^s, \quad r,s=1,2,3. \tag{5}$$

where $\{y^r\}$, r=1,2,3, is called the usual (Newtonian) velocity-coordinates. This velocity space is an affine space. As a result, we can find the so-called double-primed velocity-coordinates $\{y"^r\}$, r=1,2,3, in which the velocity space has the Euclidean structure,

$$dY^2 = \delta_{rs} dy"^r dy"^s, \quad r,s=1,2,3. \tag{6}$$

The double-primed velocity-coordinates are connected to the usual velocity-coordinates by

$$dy"^r = A^r_s(x) dy^s, \quad r,s=1,2,3, \tag{7}$$

and

$$y"^r = A^r_s(x) y^s, \quad r,s=1,2,3, \tag{8}$$

where

$$A^r_s(x) = \beta\{\delta^r_s + (\beta-1)\frac{x^r x^s}{r^2}\} \tag{9}$$

because

$$\delta_{rs} A^r_p(x) A^s_q(x) = \beta^2 g_{pq}(x). \tag{10}$$

When and only when M=0, $A^r_s(x)$ becomes $\delta^r_s$, in other words, the double-primed velocity-coordinates coincide with the usual velocity-coordinates.

Using Eq.(8), we have

$$dy"^1 dy"^2 dy"^3 = \beta^4 dy^1 dy^2 dy^3. \tag{11}$$

Eq.(6) implies

$$Y^2 = \delta_{rs} y"^r y"^s, \quad r,s=1,2,3. \tag{12}$$

Using Eq.(12), we have

$$(y")^2 = \beta^2 \{(y)^2 + \frac{r_s/r}{1-r_s/r}(\frac{1}{r}\delta_{rs}x^r y^s)^2\}, \qquad (13)$$

where $(y")^2 = \delta_{rs}y"^r y"^s$, $(y)^2 = \delta_{rs}y^r y^s$, $r,s=1,2,3$.

**Equilibrium velocity distribution in spherically symmetric gravitational field.** The Euclidean structure of the velocity space in the double-primed velocity-coordinates convinces us of the Maxwellian velocity distribution for low-energy particles [6],

$$P(y"^1,y"^2,y"^3)dy"^1 dy"^2 dy"^3 = N(\frac{m}{2\pi K_B T})^{3/2} \exp[-\frac{m}{2K_B T}(y")^2]dy"^1 dy"^2 dy"^3, \qquad (14)$$

where N is the number of particles, m their rest mass, T the temperature, and $K_B$ the Boltzmann constant. For the equilibrium velocity distribution in the spherically symmetric gravitational field of mass M represented in the usual velocity-coordinates, we put Eqs.(11) and (13) in Eq.(14) and obtain

$$P(y^1,y^2,y^3)dy^1 dy^2 dy^3 =$$
$$N\frac{(m/2\pi K_B T)^{3/2}}{(1-r_s/r)^2}\exp\{-\frac{m}{2K_B T(1-r_s/r)}[(y)^2 + \frac{r_s/r}{1-r_s/r}(\frac{1}{r}\delta_{rs}x^r y^s)^2]\}dy^1 dy^2 dy^3.$$
$$(15)$$

In the case of the spherically symmetric gravitational field, not loosing generality, we can set $x^1=r$ and $x^2=x^3=0$, and write the velocity distribution as

$$P(y^1,y^2,y^3)dy^1 dy^2 dy^3 =$$
$$N\frac{(m/2\pi K_B T)^{3/2}}{(1-r_s/r)^2}\exp\{-\frac{m}{2K_B T(1-r_s/r)}[\frac{1}{1-r_s/r}(y^1)^2 + (y^2)^2 + (y^3)^2]\}dy^1 dy^2 dy^3.$$

If we use the cylindrical coordinates $(y^1, y_h, \alpha)$, where $y^2 = y_h\cos\alpha$, $y^3 = y_h\sin\alpha$, $\alpha \in (0,2\pi)$, we can further write it as

$$P(y^1, y_h)dy^1 dy_h =$$
$$N\frac{(m/2\pi K_B T)^{3/2}}{(1-r_s/r)^2} 2\pi y_h \exp\{-\frac{m}{2K_B T(1-r_s/r)}[\frac{1}{1-r_s/r}(y^1)^2 + (y_h)^2]\}dy^1 dy_h,$$

where $y_h$ is the horizontal component of the Newtonian velocity $y^r$, $r=1,2,3$, $y_h \in (0, \infty)$, $y^1 \in (-\infty, \infty)$. We notice that the distribution function is a function of square of $y^1$, and then write the velocity distribution in the form of

$$P(y_v, y_h)dy_v dy_h =$$
$$N\frac{(m/2\pi K_B T)^{3/2}}{(1-r_s/r)^2} 4\pi y_h \exp\{-\frac{m}{2K_B T(1-r_s/r)}[\frac{1}{1-r_s/r}(y_v)^2 + (y_h)^2]\}dy_v dy_h,$$
$$(16)$$

where $y_v$ is the vertical component of the Newtonian velocity, $y_v \in (0, \infty)$.

**A test of Einstein's theory of gravitation**. The spherically symmetric gravitational field spoils the symmetry of the velocity distribution. When and only when M=0, Eq.(15) reduces to the Maxwellian distribution which is isotropic and Eq.(16) leads to the Maxwellian velocity rate distribution. In the spherically symmetric gravitational field, the pattern of the velocity distribution in the vertical direction is different from that in the horizontal direction. Moreover, for a fixed direction, it varies with the location where these low-energy particles are. We expect to examine these effects firstly on the surface of the Earth by putting the mass of the Earth, $M_\oplus$, and the radius of the Earth, $r_\oplus$, in Eq.(16), though they are small.

For high-energy particles, we have to consider some corrections from the Finslerian structures of space and time in the usual inertial coordinate system [6-8].

ACKNOWLEDGMENT

The author greatly appreciates the teachings of Prof. Wo-Te Shen.

REFERENCES

[1] A. Einstein: The Meanings of Relativity, Princeton University Press (Princeton, 1946)

[2] J. Mehra: Einstein, Hilbert and the Theory of Gravitation, D. Reidel Publishing Company (Dordrecht, 1974)

[3] Clifford M. Will, Living Review of Relativity, 4, 4 (2001) [gr-qc/0103036]

[4] Moshe Carmeri, Classical Fields: General Relativity and Gauge Theory, John Wiley & Sons, Inc. (New York, 1982)

[5] A. Einstein, Autobiographical Notes, in: A. Einstein: Philosopheo-Scientist, ed. P. A. Schipp, 3rd edition, Tudor (New York, 1970)

[6] Jian-Miin Liu, cond-mat/0108356

[7] Jian-Miin Liu, Chaos, Solitons & Fractals, 12, 399 (2001) [physics/0108044]

[8] Jian-Miin Liu, Chaos, Solitons & Fractals, 12, 1111-1135 (2001)